\documentclass[%
 reprint,
 amsmath,amssymb,
 aps
]{revtex4-2}

\usepackage{graphicx}
\usepackage{dcolumn}
\usepackage{bm}

\usepackage{tikz}

\usepackage[colorlinks=true,linkcolor=blue,urlcolor=blue,citecolor=blue]{hyperref}

\newcommand{\bsla}{\boldsymbol{\lambda}}
\newcommand{\kT}{k_{\rm B}T}
\newcommand{\md}{{\rm d}}
\newcommand{\mc}[1]{\mathcal{#1}}
\newcommand{\bs}[1]{\boldsymbol{#1}}
\newcommand{\x}{\boldsymbol{x}}
\newcommand{\W}[1]{R_{#1}}
\newcommand{\V}[1]{V_{#1}}

\begin{document}

\preprint{APS/123-QED}

\title{Hidden energy flows in strongly coupled nonequilibrium systems}

\author{Steven J.\ Large}
 \email[Current address: Viewpoint Investment Partners, Calgary, AB, T2P 4H2, Canada; ]{slarge@viewpointgroup.ca}
\author{David A.\ Sivak}%
 \email{dsivak@sfu.ca}
\affiliation{Department of Physics, Simon Fraser University, Burnaby, BC, V5A 1S6 Canada}%
 
\date{\today}

\begin{abstract}
Quantifying the flow of energy within and through fluctuating nanoscale systems poses a significant challenge to understanding microscopic biological machines. A common approach involves coarse-graining, which allows a simplified description of such systems. This has the side effect of inducing so-called hidden contributions (due to sub-resolution dynamics) that complicate the resulting thermodynamics. Here we develop a thermodynamically consistent theory describing the nonequilibrium excess power internal to autonomous systems, and introduce a phenomenological framework to quantify the hidden excess power associated with their operation. We confirm our theoretical predictions in numerical simulations of a minimal model for both a molecular transport motor and a rotary motor.
\end{abstract}

\maketitle

\section{\label{sec:introduction}Introduction}

The thermodynamic description of strongly fluctuating nanoscale systems has provided a mathematical framework in which to address some of the most fundamental physical aspects of microscopic biological systems~\cite{seifert_2012}. Building upon general results like the fluctuation theorems~\cite{jarzynski_1997,crooks_1999,kurchan_1998,lebowitz_1999}, the ability to quantify dissipation in stochastic processes out of equilibrium has uncovered a number of theoretical results limiting the achievable performance of nonequilibrium systems. For instance, the thermodynamic uncertainty relation~\cite{barato_2015,gingrich_2016} and its generalizations~\cite{pietzonka_2017,koyuk_2019,horowitz_2017} provide insight on many seemingly universal constraints faced by nonequilibrium systems~\cite{barato_2017,pietzonka_2018,pietzonka_2016,ouldridge_2017,hwang_2018,goldt_2017}.

Fundamentally, the consistency of stochastic thermodynamics relies on an assumption of timescale separation, allowing one to clearly distinguish the system from its surroundings~\cite{esposito_2012,zwanzig_book,kubo_book}. Furthermore, in its most common forms, stochastic thermodynamics also assumes weak coupling between the system of interest and the environment. In microscopic biological systems, however, both assumptions are often violated. The separation of timescales becomes blurred as there are often a number of comparable timescales relevant to a given problem, and biomolecular systems often consist of multiple strongly interacting subsystems.

For instance, the molecular machine ${\rm F}_{\rm o}{\rm F}_1$-ATP synthase (producing the majority of the cellular energy currency ATP) exhibits strong coupling between the passage of protons through the mitochondrial membrane, the mechanical rotation of a central crankshaft, and the production of ATP~\cite{soga_2017}. In fact, the chemical and mechanical aspects of the motor are inseparable from one another, both being necessary to understand the dynamics of the machine~\cite{xing_2005}. This strong coupling between chemical and mechanical processes is ubiquitous among molecular machines~\cite{brown_2020,wang_2002}.

In practice, the description of such systems is often simplified through the use of a coarse-graining procedure, whereby the state space of the full mechanochemical system is projected onto a smaller set of \emph{mesostates}, each aggregating several (unresolvable) microstates~\cite{esposito_2012}. Here, the observed dynamics of a molecular machine consists of large jumps, interleaved with small-scale fluctuations~\cite{toyabe_2011}. The large jumps are often taken to indicate a chemical reaction---producing an instantaneous change in the energy potential experienced by the molecular machine---and the full dynamics can then be mapped onto a discrete-state Markov jump model~\cite{altaner_2015,seifert_2012}. The observed chemical rates are then used to understand the thermodynamic properties (and functional capabilities) of the mechanochemical machine.  However, such an approach by its very nature ignores so-called \emph{hidden} contributions to thermodynamic quantities---such as internal energy flows---arising due to the mechanical dynamics at sub-mesostate resolutions~\cite{seifert_2019}. Such internal energy flows play an important role in understanding the interactions between the subcomponents of the molecular machine, and have been used to aid in identifying reaction coordinates in biomolecular dynamics~\cite{li_2016}.

Independent of such models, much research seeks to understand energy flows into nonequilibrium systems from a time-dependent external perturbation~\cite{schmiedl_2007,aurell_2011,sivak_2012}. These efforts have largely been restricted to deterministic driving protocols---like those typically seen in single-molecule experiments~\cite{lucero_2019,sivak_2016,zulkowski_2012,zulkowski_2014,zulkowski_2015,rotskoff_2015}---but have recently been generalized to more closely parallel the \emph{in vivo} dynamics of molecular machines~\cite{bryant_2020,large_2019,large_2018,machta_2015}. However, as of yet, none of these control-theoretical approaches to quantifying energy flows within nanoscale nonequilibrium systems have fully appreciated the essential inter-system feedback that is necessary in fully autonomous systems~\cite{large_2020,hartich_2014,horowitz_2014}.

In this article, we develop a thermodynamically complete phenomenological method for quantifying the hidden excess power \emph{within} strongly coupled nonequilibrium systems. The hidden excess power represents an energy flow communicated between the components of a strongly coupled system, and---unconstrained by the usual form of the second law---can become negative~\cite{large_2020}. Such negative excess power is a signature of a fully autonomous Maxwell demon, achieving net flow of heat from the reservoir into a subsystem---where it is transduced into work---as a result of its strong coupling with another driven subsystem.

We find that the hidden excess work per chemical transition can be decomposed into two contributions, the timescale-separated (TSS) excess work~\eqref{eq:tss-work}---the contribution to the excess work in the asymptotic TSS limit---and the nonequilibrium excess work, which is the additional contribution to the excess work when the mechanical states remain out of equilibrium at steady state. The hidden excess power (the excess work per chemical transition, averaged over the coarse-grained dynamics) is typically not easily calculable, as it requires information about hidden states. However, we provide a leading-order approximation of both the TSS excess work~\eqref{eq:tss-work-flexion-2} and nonequilibrium excess work~\eqref{eq:neq-work} which require only minimal information about the hidden-state dynamics. We demonstrate the utility of these approximations in two model systems, representing minimal models of linear and rotary mechanochemical molecular machines.

\section{\label{sec:coarse-graining}Coarse-grained representations of mechanochemical systems}

Molecular machines often couple mechanical motion to chemical reactions, for instance in kinesin~\cite{valentine_2006} and ${\rm F}_1$-ATP synthase~\cite{kawaguchi_2014}. Thus their dynamics can be described by two coupled coordinates $\x$ and $\bsla$, representing the state of mechanical and chemical subsystems~\cite{xing_2005,wang_2002}. Each chemical state $\bsla$ induces a particular potential-energy landscape $E(\x|\bsla)$ on the mechanical subsystem such that, in the absence of any chemical changes, the mechanical subsystem relaxes to a conditional equilibrium distribution
\begin{equation}
	\pi(\x|\bsla) = e^{-\beta E(\x|\bsla) + \beta F(\bsla)} \label{eq:boltzmann} \ ,
\end{equation}
with $\beta \equiv (\kT)^{-1}$ the inverse temperature of the heat bath and $F(\bsla)\equiv-\ln\int\exp\left\{ -\beta E(\x|\bsla) \right\}\md x$ the conditional equilibrium free energy given the fixed chemical state $\bsla$. 

The chemical and mechanical subsystems evolve stochastically, with joint transition $(\bsla_i,\x')\to(\bsla_j,\x)$ occurring at rate $\W{ji}^{\x,\x'}$. We further assume that the joint dynamics are bipartite, so that no simultaneous transitions in both $\x$ and $\bsla$ occur. Thus $\W{ji}^{\x,\x'} = 0$ when both $j\neq i$ and $\x\neq \x'$. Thermodynamically consistent mechanochemical dynamics---relaxing to the correct equilibrium distribution in the absence of any chemical driving (though producing nonequilibrium dynamics for nonzero chemical driving)---require that the chemical transition rates satisfy generalized detailed balance~\cite{van_den_broeck_2014,bergmann_1955},
\begin{equation}
	\frac{\W{ji}^{\x}}{\W{ij}^{\x}} = e^{-\beta\Delta G_{ji}(\x)} \ , \label{eq:detailed-balance} 
\end{equation}	
where $\W{ji}^{\x} \equiv \W{ji}^{\x,\x}$ indicates the rate of chemical transition $\bsla_i\to\bsla_j$ at fixed mechanical state $\x$, $\Delta G_{ji}(\x) = \Delta\mu_{ji} + \Delta E_{ji}(\x)$ is the change in free energy during transition $\bsla_i\to\bsla_j$, with $\Delta\mu_{ji}$ and $\Delta E_{ji}(\x) \equiv E(\x|\bsla_j) - E(\x|\bsla_i)$ the associated respective changes in chemical potential and microstate energy for the transition $\bsla_i\to\bsla_j$.

We consider transition rates obeying \eqref{eq:detailed-balance} of the general form
\begin{equation}
	\W{ji}^{\x} = \Gamma_{\rm chem} \exp\left\{-\frac{1}{2}\beta \left[\Delta\mu_{ji} + E(\x|\bsla_j) - E(\x|\bsla_i) \right]\right\} 
	\label{eq:chemical-rate} \ , 
\end{equation}
where $\Gamma_{\rm chem}$ is a kinetic prefactor (with dimensions of inverse time) that quantifies the bare rate of chemical transitions in the absence of any differences in energy or chemical potential between states.

We view the chemical dynamics as a coarse-grained representation of the mechanochemical system by defining coarse-grained states (\emph{mesostates}) that aggregate all mechanical states at a given chemical state $\bsla_i$, and their corresponding coarse-grained transition rates $\V{ji}(t) \equiv \int \W{ji}^{\x}\, p(\x,t|\bsla_i) \md\x$, for conditional probability $p(\x,t|\bsla_i)$ over mechanical states within a given mesostate~\cite{esposito_2012} 
(see SI Sec.~I)
$\V{ji}(t)$ is generally time-dependent due to the conditional microstate distribution $p(\x,t|\bsla_{i})$, but is time-independent when the mechanical dynamics are at steady state.

In the TSS limit, where at steady state the conditional distribution $p(\x|\bsla)$ of mechanical states is the conditional equilibrium distribution $\pi(\x|\bsla)$~\eqref{eq:boltzmann}, the dissipation (entropy production) of the joint system is fully determined by the coarse-grained dynamics~\cite{esposito_2012}. However, even in this limit, there are internal flows of energy and information between the mechanical and chemical subsystems~\cite{large_2020}.

\section{\label{sec:hidden-excess-work}Hidden excess work in molecular machines}

Here, we provide a method to quantify the hidden contributions to the excess power in a coarse-grained system, by viewing the hidden mechanical subsystem as being driven by the stochastic chemical dynamics.

The flow of energy between the chemical and mechanical subsystems can be quantified by the average work 
\begin{equation}
	\langle\beta W\rangle_{ji} = \beta\int \Delta E_{ji}(\x)\, p^{\rm sw}_{ji}(\x) \, \md\x
	\label{eq:switching-work}
\end{equation}
done on the mechanical subsystem by the chemical dynamics during transition $\bsla_i\to\bsla_j$. The average is over the switching-position distribution $p^{\rm sw}_{ji}(\x)$ of mechanical states $\x$ from which the chemical transition $\bsla_i\to\bsla_j$ occurs~\cite{large_2020}, which in general depends on the dynamics of both the chemical and mechanical subsystems.

The average \emph{excess work} is $\langle\beta W_{\rm ex}\rangle_{ji} \equiv \langle\beta W\rangle_{ji} - \beta\Delta F_{ji}$, for change $\Delta F_{ji} \equiv F(\bsla_j) - F(\bsla_i)$ in conditional equilibrium free energy upon the chemical transition. At steady state, the average rate of excess work (the \emph{excess power}) on the mechanical subsystem is expressed in terms of coarse-grained rates as
\begin{equation}
\langle \beta\mc{P}_{\rm ex}\rangle_{\Lambda \to X} \equiv \sum_{i,j}P_{i}\V{ji}\langle \beta W_{\rm ex}\rangle_{ji} \ . \label{eq:excess-power}
\end{equation}
$P_i$ is the probability of the mesostate with chemical state $\bsla_i$, and angle brackets $\langle\cdots\rangle_{\Lambda\to X}$ indicate an average over the dynamics of the chemical subsystem $\Lambda$ and mechanical subsystem $X$.

Motivated by the approach to discrete control protocols in Ref.~\cite{large_2019}, we decompose the average excess work for transition $\bsla_i\to\bsla_j$ into two components,
\begin{equation}
	\langle\beta W_{\rm ex}\rangle_{ji} = \langle\beta W_{\rm ex}^{\rm TSS}\rangle_{ji} + \langle\beta W_{\rm ex}^{\rm neq}\rangle_{ji} \label{eq:work-splitting} \ ,
\end{equation}
where $\langle \beta W_{\rm ex}^{\rm TSS}\rangle_{ji}$ is the excess work in the timescale-separated (TSS) limit where the mechanical subsystem fully equilibrates between chemical reactions---but still evolves slowly when compared with the environmental degrees of freedom, so that the reservoirs can still be clearly thermodynamically distinguished from the system---and $\langle\beta W_{\rm ex}^{\rm neq}\rangle_{ji}$ is the nonequilibrium excess work quantifying the additional work beyond the TSS work due to the mechanical subsystem being out of equilibrium.

\subsection{\label{subsec:tss-work}TSS excess work}

For the transition rates in \eqref{eq:chemical-rate}, the timescale-separated switching-position distribution is the normalized geometric mean of the conditional equilibrium distributions at $\bsla_i$ and $\bsla_j$,
\begin{equation}
	p^{\rm sw}_{ji}(\x) = 
	\frac{\sqrt{\pi_i(\x)\pi_j(\x)}}{\mc{Z}^{\rm sw}_{ji}} 
	\label{eq:switching-position-distribution} \ ,
\end{equation}
where $\pi_i(\x) \equiv \pi(\x|\bsla_i)$, and $\mc{Z}^{\rm sw}_{ji} \equiv \int\sqrt{\pi_i(\x)\pi_j(\x)}\, \md{\x}$ normalizes the distribution. In information theory, $\mc{Z}^{\rm sw}_{ji}$ is the \emph{Bhattacharyya coefficient}~\cite{bhattacharyya_1943}, a measure of the difference between two probability distributions. Inserting \eqref{eq:switching-position-distribution} into \eqref{eq:switching-work} 
and subtracting the free energy difference $\Delta F_{ji}$ gives an exact relation for the TSS excess work 
(see SI Sec.~II)
\begin{subequations}
\label{eq:tss-work}
\begin{align}	
	\langle\beta W_{\rm ex}^{\rm TSS}\rangle_{ji} &= \frac{1}{\mc{Z}^{\rm sw}_{ji}}\int \sqrt{\pi_i(\x)\pi_j(\x)} \, \ln\frac{\pi_i(\x)}{\pi_j(\x)}\, \md\x \label{eq:tss-work-1} \\ 
	&= D\left( \sqrt{\pi_i\pi_j} \, || \, \pi_j \right) - D\left( \sqrt{\pi_i\pi_j} \, || \, \pi_i \right) \ , \label{eq:tss-work-2}
\end{align}
\end{subequations}
for \emph{relative entropy} $D(p||q) \equiv \int p(\x)\ln\frac{p(\x)}{q(\x)}\md\x$ between two distributions~\cite{cover-thomas}. Intuitively, the TSS excess work quantifies the increase in additional free energy of the switching-position distribution $p^{\rm sw}_{ji}(\x)$ through the chemical reaction $\bsla_i\to\bsla_j$~\cite{sivak_2012_FE}. (Eq.~\eqref{eq:tss-work-2} was also derived for master-equation dynamics for a general switching-position distribution in \cite{large_2020}.)

The TSS excess work is antisymmetric under index exchange, $\langle\beta W_{\rm ex}^{\rm TSS}\rangle_{ji} = - \langle\beta W_{\rm ex}^{\rm TSS}\rangle_{ij}$, and thus the TSS excess power~\eqref{eq:excess-power} (obtained by averaging the transition-specific excess work over coarse-grained dynamics) is 
\begin{equation}
	\langle\beta \mc{P}_{\rm ex}^{\rm TSS}\rangle_{\Lambda\to X} = \sum_{i < j} \left( P_i \V{ji} - P_j \V{ij} \right)\langle\beta W_{\rm ex}^{\rm TSS}\rangle_{ji} \label{eq:tss-power-total} \ .
\end{equation}
Thus, even for nonzero $\langle\beta W_{\rm ex}^{\rm TSS}\rangle_{ji}$ for chemical transition $\bsla_i \to \bsla_j$, the associated hidden excess power averaged over transitions in each direction vanishes when there is no net chemical flux ($P_i V_{ji} = P_j V_{ij}$). This occurs, for instance, when there is no chemical driving ($\Delta\mu = 0$), and is a direct result of the autonomous nature of the system, exemplifying the essential importance of requiring the rates~\eqref{eq:chemical-rate} to satisfy local detailed balance~\eqref{eq:detailed-balance}. 

The TSS excess work further simplifies in the small-$\Delta\bsla$ limit, where Taylor expanding \eqref{eq:tss-work-1} produces 
(see SI Sec.~III)
\begin{equation}
	\langle\beta W_{\rm ex}^{\rm TSS}\rangle_{ji} \approx \frac{1}{4!}\mc{S}_{m\ell k}(\bsla_i)\Delta\lambda_{ji}^{m}\Delta\lambda_{ji}^{\ell}\Delta\lambda_{ji}^k \label{eq:tss-work-flexion-1} \ ,
\end{equation}
where $\Delta\lambda_{ji}^k$ is the change in the $k$th component of the chemical state vector $\bsla$ during the transition $\bsla_i\to\bsla_j$, and the rank-3 tensor
\begin{equation}
    \mc{S}_{m\ell k}(\bsla) = \langle\delta f_m\delta f_{\ell}\delta f_k\rangle_{\bsla_i}\label{eq:tss-work-flexion-2} \ ,
\end{equation} 
is the third centered moment of conjugate forces $f_k \equiv -\partial_{\lambda^k}E(\x|\bsla)$. Here, we use an Einstein summation notation, where repeated indices are implicitly summed over. In contrast, the analogous infinite-time excess work in \cite{large_2019} is second order in $\Delta\bsla$ and uses the force variance $\langle\delta f_m\delta f_k\rangle_{\bsla_i}$ instead of the third centered moment. Here, the feedback of the mechanical subsystem on the chemical dynamics results in cancellation of the second-order terms.

\subsection{\label{subsec:neq-work}Nonequilibrium excess work}

In the small-$\Delta\bsla$ limit, linear-response theory provides a simple approximation for the nonequilibrium excess work $\langle\beta W_{\rm ex}^{\rm neq}\rangle_{ji}$. In particular, if upon the chemical transition $\bsla_i\to\bsla_j$, the corresponding change in the mechanical energy landscape $E(\x|\bsla_i)\to E(\x|\bsla_j)$ is small, $E(\x|\bsla_j)$ is well approximated by a first-order Taylor expansion about $\bsla_i$.  We additionally assume a moderate timescale separation between the mechanical and chemical dynamics, such that the relaxation time of the conjugate forces $f_k$ is significantly shorter than the chemical-state dwell time. This constraint is weaker than the timescale-separation limit typically considered in coarse-graining, where the hidden states are assumed to relax infinitely faster than the chemical dynamics and thus equilibrate fully~\cite{esposito_2012}.  

Under these assumptions, the leading-order contribution to the nonequilibrium excess work during chemical transition $\bsla_i\to\bsla_j$ is 
(see SI Sec.~IV)
\begin{equation}
	\langle\beta W_{\rm ex}^{\rm neq}\rangle_{ji} \approx \beta\Gamma_{ji} \zeta_{kk'}(\bsla_i)\Delta\lambda_{ji}^k\frac{\sum_{s}P_s\V{is}\Delta\lambda_{is}^{k'}}{V_{i*}} \label{eq:neq-work} \ .
\end{equation}
$\Gamma_{ji} \equiv \Gamma_{\rm chem} \exp\left\{-\tfrac{1}{2}\beta\Delta\mu_{ji}\right\}$, $V_{i*} \equiv \sum_{s}P_s\V{is}$ is the total rate of chemical transitions entering $\bsla_i$, and the summation index $s$ runs over all chemical states. $\zeta_{kk'}(\bsla) \equiv \beta\int_0^{\infty}\langle\delta f_k(0)\delta f_{k'}(t)\rangle_{\bsla}\, \md t$ is the \emph{generalized friction tensor}~\cite{sivak_2012} that quantifies the leading-order dissipation due to nonequilibrium driving by a deterministic control protocol. Equation~\eqref{eq:neq-work} quantifies the nonequilibrium excess work using information about the coarse-grained dynamics and the friction tensor $\zeta_{kk'}(\bsla)$, which is determined from the hidden states' equilibrium fluctuations.

At steady state, the excess power by the chemical dynamics on the mechanical subsystem averages over the coarse-grained transitions,
\begin{equation}
	\langle \beta\mc{P}_{\rm ex}\rangle_{\Lambda\to X} = \sum_{i,j}\V{ji}P_i\left[ \langle\beta W_{\rm ex}^{\rm TSS}\rangle_{ji} + \langle \beta W_{\rm ex}^{\rm neq}\rangle_{ji} \right] \ .
\end{equation}
The near-equilibrium expressions for the TSS~\eqref{eq:tss-work} and nonequilibrium excess work~\eqref{eq:neq-work} in terms of conditional equilibrium averages permits estimation of the excess power in experimental investigation of autonomous mechanochemical molecular machines: the low-resolution observations of hidden-state fluctuations it requires are more tractable than inferring the full mechanochemical dynamics of the motor~\cite{ariga_2018,toyabe_2010}.

\section{\label{sec:model-systems}Model systems}

\begin{figure}
	\centering\includegraphics[width=\columnwidth]{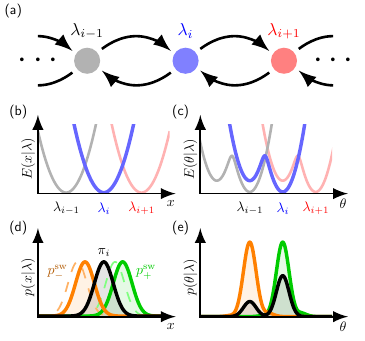}
	\caption{\label{fig:schematic}{\bf Schematic of model mechanochemical machines.}
	(a) Chemical reaction diagram, an infinite one-dimensional lattice of chemical states.
	(b-c) Conditional energy landscapes for (b) linear $E(x|\lambda)$ and (c) rotary $E(\theta|\lambda)$ models. 
	(d-e) Equilibrium distribution (black) and forward/reverse equilibrium switching-position distributions (solid green/orange) for $\lambda_i$ (blue state in (b-c)) for (d) linear and (e) rotary models.
    Dashed curves in (d) indicate nonequilibrium switching-position distributions for net chemical
    flux towards higher-index $\lambda$ (to the right in (a)).}
\end{figure}

We illustrate our theoretical predictions by investigating the hidden excess power in two model systems, representing linear-transport and rotary molecular motors. In both cases, the chemical coordinate evolves on a discrete, one-dimensional lattice, with only nearest-neighbor transitions having rates given by \eqref{eq:chemical-rate}. The mechanical coordinate $x$ ($\theta$ in the rotary model) diffuses in a one-dimensional energy landscape $E(x|\lambda)$ ($E(\theta|\lambda)$) determined by the instantaneous chemical state $\lambda$. Figure~\ref{fig:schematic} shows a schematic of both model systems.

In the linear-transport motor, the TSS excess work~\eqref{eq:tss-work} is zero, thus serving as a model to isolate the nonequilibrium excess work~\eqref{eq:neq-work}. Conversely, for the rotary motor we analyze its excess work in the TSS limit where the nonequilibrium excess work is zero, thereby isolating the effects of the TSS excess work~\eqref{eq:tss-work}.

\subsection{\label{subsec:linear-system}Linear-transport motor}

The linear-transport motor has a one-dimensional mechanical degree of freedom subject to a harmonic potential $E(x|\lambda_i) = \tfrac{1}{2}k_{\rm trap}(x - \lambda_i)^2$, with the instantaneous chemical state $\lambda_i$ determining the potential minimum. Figure~\ref{fig:schematic}a-b shows a schematic.

\begin{figure}
	\centering\includegraphics[width=\columnwidth]{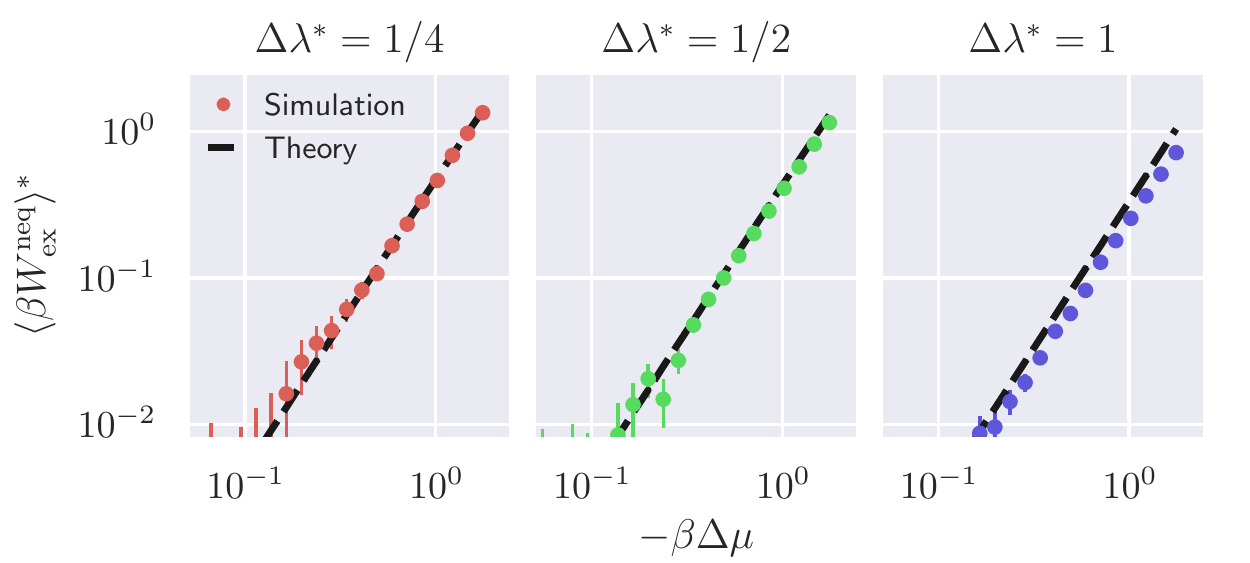}
	\caption{\label{fig:harmonic-work}{\bf Nonequilibrium excess work for linear-transport motor.} Nonequilibrium excess work $\langle\beta W_{\rm ex}^{\rm neq}\rangle^* \equiv \langle\beta W_{\rm ex}^{\rm neq}\rangle / (\beta\Gamma_{\rm chem}\zeta\Delta\lambda^2)$, as a function of forward chemical-potential bias $-\beta\Delta\mu$, for numerical simulation (dots) or approximate theory~\eqref{eq:neq-work-1D} (black dashed line). Different panels show different step sizes, non-dimensionalized by the 
	equilibrium standard deviation $\sigma_x = 1/\sqrt{\beta k_{\rm trap}}$ of the 
	mechanical coordinate at fixed $\lambda$: $\Delta\lambda^* \equiv \Delta\lambda/\sigma_x$.}
\end{figure}

The equilibrium mechanical distributions for consecutive chemical states are only distinguished by their mean $\lambda_i$, so we observe the system in a comoving frame by changing mechanical coordinate to the relative position $x-\lambda_i$.  This reference frame has a steady-state distribution for time-independent transition rates~\eqref{eq:chemical-rate} and uniform fixed $\Delta\mu_{i+1,i}=\Delta\mu<0$ biasing the machine to (on average) move forward. Here, the generalized friction is uniform and equals the viscous friction, $\zeta=\gamma$, and the steady-state coarse-grained forward and reverse rates $V_{\pm}$ and probabilities $P_i$ are uniform. Equation~\eqref{eq:neq-work} approximates the steady-state nonequilibrium excess work, simplifying in this case to 
(see SI Sec.~IV)
\begin{equation}
	\langle\beta W_{\rm ex}^{\rm neq}\rangle_{\Delta\lambda} \approx \beta\zeta\Delta\lambda^2 \frac{\left( V_+ - V_- \right) \left( V_+\Gamma_{+} - V_-\Gamma_- \right)}{\left( V_+ + V_- \right)^2} \label{eq:neq-work-1D} \ .
\end{equation} 
Here, the nonequilibrium excess work depends on the chemical driving $\Delta\mu$ through both the coarse-grained chemical rates $V_{\pm}\propto\exp\left\{\pm\tfrac{1}{2}\beta\Delta\mu\right\}$ (following from their dependence on microscopic transition rates $R_{ji}^x$ and \eqref{eq:chemical-rate}) and the energy-independent rate $\Gamma_{\pm}\propto\exp\left\{\pm\tfrac{1}{2}\beta\Delta\mu\right\}$. Figure~\ref{fig:harmonic-work} compares the theoretical predictions~\eqref{eq:neq-work-1D} with numerical results, showing good agreement in the small-$\Delta\lambda$ limit.

\subsection{\label{subsec:rotary-system}Rotary motor}
Next, we consider the Kawaguchi-Sasa-Sagawa (KSS) model of ${\rm F}_1$-ATPase~\cite{kawaguchi_2014}. Here, the rotational angle $\theta$ of ${\rm F}_1$-ATPase's crankshaft evolves in the potential of mean force (PMF) 
\begin{align} 
    \beta E_{\rm KSS}(\theta|\lambda_i) &= \tfrac{1}{2}\beta k_{\rm c}\left( \theta - \lambda_i \right)^2 \label{eq:energy-kss} \\
	&\quad -\ln\left[ e^{-\beta k_{\rm c}\phi(\theta - \lambda_i)} + e^{\beta\Delta E_{\rm S} + \tfrac{1}{2}\beta k_{\rm c}\phi^2} \right]  \ ,  \nonumber
\end{align}
arising from fast switching between two harmonic potentials, each with stiffness $k_{\rm c}$, with angular offset $\phi$ and energetic offset $\Delta E_{\rm S}$. Figure~\ref{fig:schematic}a,c shows a schematic. Unlike the (purely harmonic) linear-transport model 
(in the previous subsection), 
for $\Delta E_{\rm S} \neq 0$ and $\phi\neq 0$ this potential is asymmetric in $\theta$, generally producing non-zero TSS excess work~\eqref{eq:tss-work-2}. To isolate the effects of the nonzero TSS excess work, we consider this model in the TSS limit, where the mechanical subsystem comes to conditional equilibrium between each chemical transition, and the nonequilibrium excess work~\eqref{eq:neq-work} vanishes.

\begin{figure}
\centering
\includegraphics[width=\columnwidth]{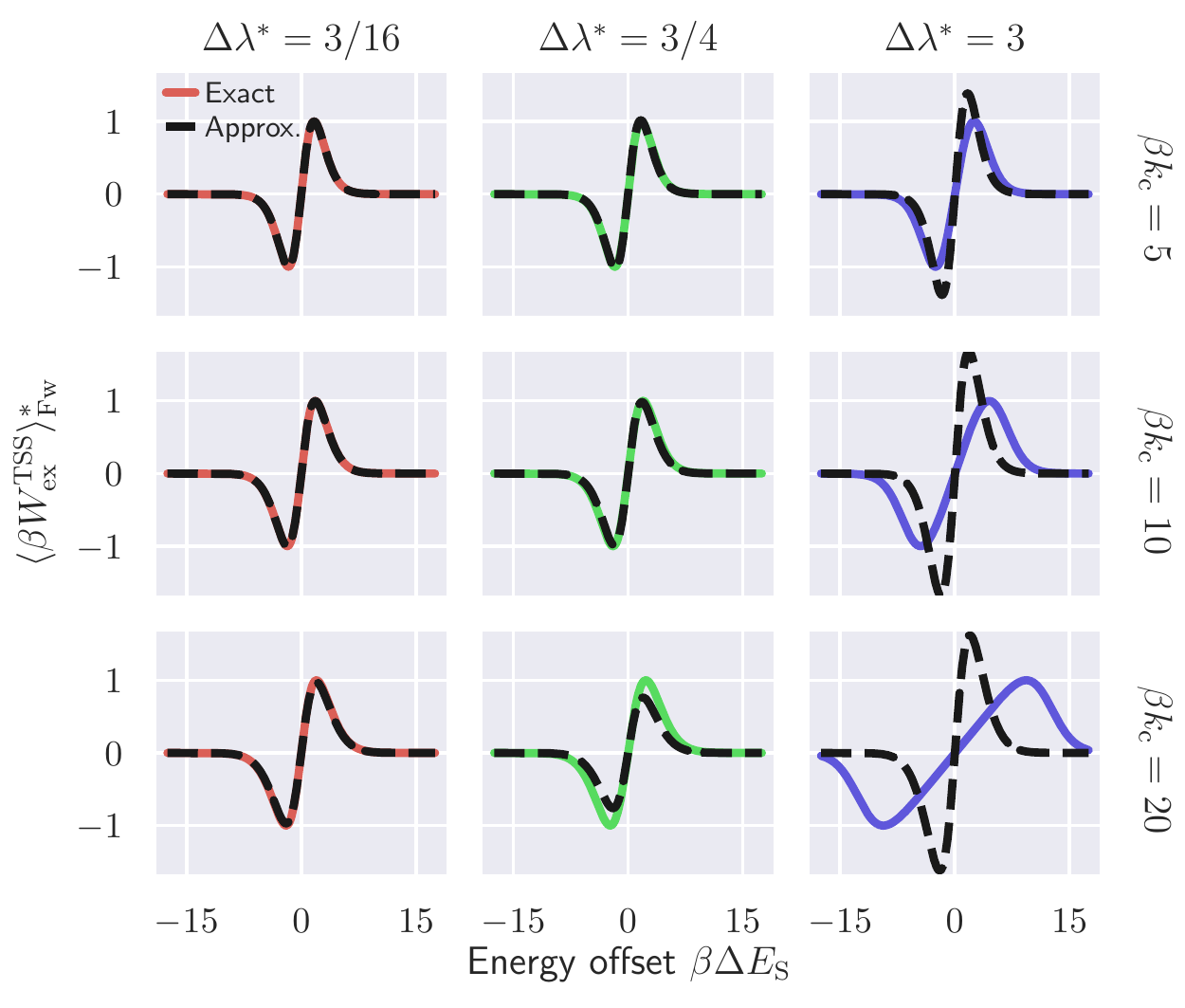}
\caption{\label{fig:kss-work}
{\bf TSS excess work for a rotary motor.}
TSS work $\langle\beta W_{\rm ex}^{\rm TSS}\rangle_{\rm Fw}^* \equiv \langle\beta W_{\rm ex}^{\rm TSS}\rangle_{\rm Fw}/\max_{\Delta E_{\rm S}}\langle\beta W_{\rm ex}^{\rm TSS}\rangle_{\rm Fw}$ per forward step, as a function of energy offset $\Delta E_{\rm S}$ (nondimensionalized by the angular offset between minima), for several chemical step sizes $\Delta\lambda^*\equiv \Delta\lambda/\phi$ (columns) and spring constants $\beta k_{\rm c}$ (rows). 
Colored solid curves: exact; dashed curves: small-$\Delta\lambda$ approximation~\eqref{eq:tss-work-flexion-1}.} 
\end{figure}

Figure~\ref{fig:kss-work} compares the exact TSS excess work $\langle\beta W_{\rm ex}^{\rm TSS}\rangle_{\rm Fw}$ per forward step~\eqref{eq:tss-work} with its leading-order approximation~\eqref{eq:tss-work-flexion-1}, as a function of the energy offset $\Delta E_{\rm S}$. For small (but nonzero) energy offset $\Delta E_{\rm S}$, the switching-position distribution~\eqref{eq:switching-position-distribution} for forward steps is heavily concentrated in the right minimum of the potential (Fig.~\ref{fig:schematic}e), where for positive (negative) $\Delta E_{\rm S}$ the changes in energy of the mechanical state $\theta$ upon the chemical transition are positive (negative). For large-magnitude offsets $|\Delta E_{\rm S}|$, the potential becomes effectively harmonic, and thus the TSS work~\eqref{eq:tss-work} vanishes. For smaller $\Delta\lambda/\phi$ and lower $\beta k_{\rm c}$, the TSS excess work approximation~\eqref{eq:tss-work-flexion-1} agrees well with the exact calculation~\eqref{eq:tss-work}. For large step sizes and stronger spring constants, the approximation in \eqref{eq:tss-work-flexion-1} begins to break down. 

At the level of the (driven) mechanical subsystem, negative TSS excess work indicates net heat flow from the thermal reservoir into the mechanical subsystem and onward as work extracted by the chemical dynamics. The strong coupling between mechanical and chemical subsystems ensures thermodynamic consistency by a compensating heat flow from the chemical subsystem to the thermal reservoir that rescues what would otherwise be a violation of the second law~\cite{horowitz_2014,hartich_2014}. In such circumstances, the chemical dynamics operate as a fully autonomous Maxwell demon, effectively using the information gleaned through strong coupling to the mechanical subsystem to transduce heat from the bath into extracted work.

\section{\label{sec:discussion}Discussion}

In this article, we present a phenomenological formalism to estimate the hidden excess power internal to coarse-grained autonomous nonequilibrium systems, which only requires the coarse-grained chemical dynamics and minimal information about the hidden mechanical dynamics. This theoretical framework provides a means to estimate the hidden energy flows within molecular machines without explicitly modeling their microscopic dynamics.

We identify two distinct contributions to the hidden excess work: the TSS excess work~\eqref{eq:tss-work}---which persists in the TSS limit, when between chemical reactions the hidden mechanical subsystem fully relaxes to a conditional equilibrium---and the nonequilibrium excess work~\eqref{eq:neq-work}, which is the additional energetic cost due to the mechanical subsystem being out of equilibrium. 

The exact TSS excess work $\langle\beta W_{\rm ex}^{\rm TSS}\rangle_{ji}$ for a particular chemical transition is the difference between two relative entropies~\eqref{eq:tss-work}, and can be interpreted as quantifying the concomitant change of the additional free energy in the mechanical subsystem~\cite{large_2020}. Unlike the analogous \emph{infinite-time excess work} in \cite{large_2019}, the TSS excess work for any particular transition can be either positive or negative. This is consistent with findings in Ref.~\cite{large_2020}, where we showed that---due to the strong coupling between the mechanical and chemical subsystems---the excess power $\langle\beta\mc{P}_{\rm ex}\rangle_{\Lambda\to X}$ (the rate of excess work, averaged over all coarse-grained transitions) is not, in itself, constrained by the second law, and thus can become negative.

We also identify a phenomenological expression that approximates (to leading order in $\Delta\bsla$) the TSS work of the chemical dynamics on the mechanical system by the third centered moment of the conditional equilibrium distribution of conjugate forces~\eqref{eq:tss-work-flexion-2}. This is in contrast to a similar expansion of the infinite-time excess work in \cite{large_2019} that is second-order in $\Delta\bsla$ and uses the force variance in place of the third moment in \eqref{eq:tss-work-flexion-2}. Here, feedback from the mechanical subsystem to the chemical dynamics leads to exact cancellation of terms 
that are 
second order 
in $\Delta\bsla$. Additionally, in the moderate timescale-separated limit, the nonequilibrium excess work is determined by the generalized friction tensor~\eqref{eq:neq-work}, which can be inferred through observations of equilibrium force fluctuations.

While this decomposition of the average excess work per transition is reminiscent of the adiabatic/nonadiabatic decomposition of the entropy production~\cite{esposito_2010_PRL}, the individual terms in \eqref{eq:work-splitting} bear no direct relationship with either contribution to the entropy production; moreover, the excess work internal to such autonomous systems is not directly related to entropy production~\cite{large_2020}.

Our theoretical framework can be used as a tool in experimental studies of molecular machines to better understand excess power in autonomous systems and address how internal energy flows influence whole-system function. In particular, recent experiments on kinesin~\cite{ariga_2018} and ${\rm F}_1$ ATPase~\cite{toyabe_2010} have used their mechanical fluctuations to infer hidden entropy production; our theoretical framework can make use of these measurements of mechanical fluctuations to approximate both the TSS excess work and nonequilibrium excess work. 

Furthermore, the identification of negative steady-state excess power has an interesting physical interpretation, providing a signature of Maxwell-demon behavior in the mechanochemical machine~\cite{hartich_2014}. Specifically, negative excess power means that, on average, there is net heat flow from the reservoir into the subsystem being driven. We showed an example of this with the TSS excess work in the rotary model system, and analogous behavior was recently found in a similar bistable potential in Ref.~\cite{large_2020}. Our framework provides a tractable method to measure this quantity, and can be used to identify such information-thermodynamic features in biological molecular machines.

\acknowledgments
We thank Jannik Ehrich (SFU Physics) for insightful discussions and Emma Lathouwers (SFU Physics) and Miranda Louwerse (SFU Chemistry) for comments on the manuscript. This work is supported by a Natural Sciences and Engineering Research Council of Canada (NSERC) CGS Doctoral fellowship (SJL), an NSERC Discovery Grant and Discovery Accelerator Supplement (DAS), a Tier-II Canada Research Chair (DAS), and Westgrid (www.westgrid.ca) and Compute Canada Calcul Canada (www.computecanada.ca).

%

\appendix



\begin{widetext}

\section*{Supplementary Material: Hidden excess power in autonomous coarse-grained nonequilibrium systems}

\end{widetext}

\section{\label{suppsec:entropy-production}Coarse-grained representations of mechanochemical systems}

For a mechanochemical system with discrete chemical state $\bsla$ and continuous mechanical state $\x$, the dynamics of the joint probability $p_j(\x)$ for system state $(\bsla_j,\x)$ is governed by the master equation~\cite{gardiner}
\begin{equation}
	\md_t p_j(\x) = \sum_{i}\int_{x'} \W{ji}^{\x,\x'} p_i(\x') \md\x' \ .
	\label{eq:app:master-equation}
\end{equation}
$\W{ji}^{\x,\x'}$ is the rate of transitions between the joint states $(\bsla_i,\x')\to(\bsla_j,\x)$, and the summation runs over all chemical states $\bsla_i$. Writing the joint probability as $p_j(\x') = P_j \, p(\x'|\bsla_j)$---where $P_j \equiv \int_{\bs{x}}p_j(\x)\md \x$ is the marginal distribution of chemical state $\bsla_j$---gives
\begin{equation}
	\md_t p_j(\x) = \sum_{i,j} P_i \int_{\x'} \W{ji}^{\x,\x'} p(\x'|\bsla_i)\md\x' \ .
\end{equation}
Integrating over mechanical states $\x$ gives the equation of motion for the coarse-grained mesostates,
\begin{equation}
	\md_t P_j = \sum_{i,j} P_i \int_{\x} \int_{\x'} \W{ji}^{\x,\x'} p(\x'|\bsla_i) \md\x'\md\x \ .
\end{equation}
Further constraining the joint dynamics to be bipartite (so there are no simultaneous transitions in both $\x$ and $\bsla$ and thus $\W{ji}^{\x,\x'} = 0$ if $j\neq i$ and $\x\neq \x'$), the coarse-grained master equation simplifies to
\begin{subequations}
\begin{align}
	\md_t P_j &= \sum_{i,j} P_i \int_x \W{ji}^{\x} p(\x|\bsla_i)\md\x \\
	&= \sum_{i,j}\V{ji} P_i \ ,
\end{align}
\end{subequations}
where $\W{ji}^{\x} \equiv \W{ji}^{\x,\x}$ is the bipartite transition rate at fixed mechanical state $\x$, and $\V{ji} \equiv \int_{\x} \W{ji}^{\x}p(\x|\bsla_i)\md\x$ is the coarse-grained rate, the observed transition rate of $\bsla_i\to\bsla_j$ if the mechanical states are hidden~\cite{esposito_2012}.

\section{\label{suppsec:TSS-power}Excess work in timescale-separated autonomous systems}

Here we consider the contribution to the excess work of an autonomous chemically driven mechanical subsystem in the timescale-separated (TSS) limit.  In this context, chemical transitions $\bsla_i\to\bsla_j$ induce an instantaneous change in the energy landscape $E(\x|\bsla)$ governing the dynamics of mechanical subsystem $\x$. By assumption, in the TSS limit, before each chemical transition occurs the mechanical subsystem is at equilibrium, with microstates distributed as
\begin{equation}
	\pi_i(\x) = e^{-\beta E(\x|\bsla_i) + \beta F(\bsla_i)} \label{eq:app:boltzmann} \ .
\end{equation}	
The work performed by the chemical dynamics when the particular transition $\bsla_i\to\bsla_j$ occurs is
\begin{equation}
	\langle W^{\rm TSS}\rangle_{ji} = \int \left[ E(\x|\bsla_j) - E(\x|\bsla_i) \right]p_{ji}^{\rm sw}(\x)\md \x\label{eq:app:TSS-work-1} \ ,
\end{equation}
where the angle brackets $\langle\cdots\rangle_{ji}$ indicate an average over the equilibrium switching-position distribution $p^{\rm sw}_{ji}(\x)$, namely the distribution of positions from which the chemical transition $\bsla_i\to \bsla_j$ occurs.

When the chemical dynamics are independent of the mechanical state, the switching-position distribution is simply the equilibrium distribution~\eqref{eq:app:boltzmann}, and the average TSS work reduces to the \emph{infinite-time work} from \cite{large_2019}.  In this case, the excess work associated with a discrete change in the energy landscape is the relative entropy between the adjacent equilibrium distributions (Eq.~(8) in \cite{large_2019}). However, for an autonomous system that obeys microscopic reversibility, the chemical transition rates must obey generalized detailed balance~(2), and the switching-position distribution in general differs from the equilibrium distribution.  For the rates~(3) considered in the main text, the switching-position distribution is the normalized product of the equilibrium distribution and the chemical transition rate:
\begin{equation}
	p^{\rm sw}_{ji}(\x) \equiv \frac{1}{\mc{Z}^{\rm sw}_{ji}}\pi_i(\x) e^{-\frac{1}{2}\beta\left[ E({\x}|\bsla_j) - E(\x|\bsla_i) \right]} \ ,
\end{equation}
where $\mc{Z}^{\rm sw}_{ji}$ is the normalization constant, which contains the prefactors of the rate expression~(3). Inserting the equilibrium distribution~\eqref{eq:app:boltzmann} gives
\begin{subequations}
\label{eq:app:switching-dist-1}
\begin{align}
	p^{\rm sw}_{ji}(\x) &= \frac{1}{\mc{Z}^{\rm sw}_{ji}}e^{-\beta E(\x|\bsla_i) -\frac{1}{2}\beta\left[ E(\x|\bsla_j) - E(\x|\bsla_i) \right]} \label{eq:app:switching-dist-1-1} \\
	&= \frac{1}{\mc{Z}^{\rm sw}_{ji}}e^{-\frac{1}{2}\beta\left[ E(\x|\bsla_i) + E(\x|\bsla_j) \right]} \label{appendix:switching-dist-1-2} \\
	&= \frac{1}{\mc{Z}^{\rm sw}_{ji}}\sqrt{\pi_i\pi_j} \label{eq:app:switching-dist-1-3} \ ,
\end{align}
\end{subequations}
the normalized geometric mean of the equilibrium distributions at $\bsla_i$ and $\bsla_j$. The normalization factor is
\begin{equation}
	\mc{Z}^{\rm sw}_{ji} = \int \sqrt{\pi_i\pi_j}\, \md \x \label{eq:app:BC} \ .
\end{equation}
(To simplify notation, we suppress the explicit $\x$-dependence of the equilibrium distributions $\pi_i$ throughout this appendix.)

Substituting in \eqref{eq:app:TSS-work-1} the energies in terms of the corresponding equilibrium probabilities,
\begin{equation}
	\beta E(\x|\bsla_i) = -\ln \pi_i(\x) +\beta F(\bsla_i) \label{eq:app:energy-distribution-correspondence} \ ,
\end{equation}
and the equilibrium switching-position distribution~\eqref{eq:app:switching-dist-1-3}, the TSS work becomes
\begin{equation}
	\langle\beta W^{\rm TSS}\rangle_{ji} = \frac{1}{\mc{Z}^{\rm sw}_{ji}}\int \sqrt{\pi_i\pi_j} \, \ln\frac{\pi_i}{\pi_j} \, \md \x + \beta\Delta F_{ji} \ , \label{eq:app:tss-work}
\end{equation}
where $\Delta F_{ji} \equiv F(\bsla_j) - F(\bsla_i)$ is the equilibrium free energy difference between chemical states $\bsla_i$ and $\bsla_j$.  Thus the \emph{excess work} is
\begin{subequations}
\begin{align}
	\langle\beta W_{\rm ex}^{\rm TSS}\rangle_{ji} &\equiv \langle\beta W^{\rm TSS}\rangle{ji} - \beta\Delta F_{ji} \\
	&= \frac{1}{\mc{Z}^{\rm sw}_{ji}}\int \sqrt{\pi_i\pi_j}\, \ln\frac{\pi_i}{\pi_j}\, \md \x \label{eq:app:TSS-work-2} \ .
\end{align}
\end{subequations}

The transition-specific TSS excess work also equals a difference between relative entropies:
\begin{subequations}
\label{eq:app:TSS-work-3}
\begin{align}
	\langle\beta W_{\rm ex}^{\rm TSS}\rangle_{ji} &= \frac{1}{\mc{Z}^{\rm sw}_{ji}}\int \sqrt{\pi_i\pi_j}\ln\frac{\pi_i}{\pi_j}\md \x \label{eq:app:TSS-work-3-1} \\
	&= \frac{1}{\mc{Z}^{\rm sw}_{ji}}\int\left( \ln\frac{\sqrt{\pi_i\pi_j}}{\pi_j} - \ln\frac{\sqrt{\pi_i\pi_j}}{\pi_i} \right)\sqrt{\pi_i\pi_j} \, \md \x \label{eq:app:TSS-work-3-2} \\
	&= D\left( \sqrt{\pi_i\pi_j} \, || \, \pi_j \right) - D\left( \sqrt{\pi_i\pi_j} \, || \, \pi_i \right) \label{eq:app:TSS-work-3-3} \ .
\end{align}
\end{subequations}

\section{\label{suppsec:TSS-power-expansion}Expansion of the TSS work}

While the expressions for the TSS excess work in (8) are exact, it is convenient to examine limiting cases so as to approximate the excess work in terms of equilibrium averages.  For instance, the expansion of the relative entropy in \cite{large_2019} shows that, for small $\Delta\bsla$, the conjugate-force variance can be used to approximate the excess work associated with a discrete transition, when the discrete transitions are independent of the mechanical state $\x$.  Here, we proceed similarly, Taylor expanding the integrand of (8a) in $\Delta\bsla$. To start we define 
\begin{equation}
	g(\x,\bsla_i,\bsla_j) \equiv \sqrt{\pi_i\pi_j}\ln\frac{\pi_i}{\pi_j} \label{eq:app:TSS-work-expansion-1} \ ,
\end{equation}
the integrand of (8a), where we omit the explicit $\x$-dependence of the equilibrium distribution $\pi_i$. For small steps $\Delta\bsla_{ji} = \bsla_j - \bsla_i$, we Taylor expand \eqref{eq:app:TSS-work-expansion-1} in $\bsla_j$ about $\bsla_i$,
\begin{align}
	g(&\x,\bsla_i,\bsla_j) \approx g(\x,\bsla_i,\bsla_i) + \left[ \partial_{\lambda_j^k}g(\x,\bsla_i,\bsla_j) \right]_{\lambda_i^k}\Delta\lambda^k_{ji}\nonumber \\
	&+ \frac{1}{2}\left[ \partial^2_{\lambda_j^{\ell}\lambda_j^k}g(\x,\bsla_i\bsla_j) \right]_{\lambda_i^k\lambda_i^{\ell}}\Delta\lambda_{ji}^k\Delta\lambda_{ji}^{\ell} 	\label{eq:app:TSS-work-expansion-2} \\
	&+ \frac{1}{3!}\left[ \partial^3_{\lambda_j^m\lambda_j^{\ell}\lambda_j^k}g(\x,\bsla_i,\bsla_j) \right]_{\lambda_i^k\lambda_i^{\ell}\lambda_i^{m}}\Delta\lambda_{ji}^k\Delta\lambda_{ji}^{\ell}\Delta\lambda_{ji}^{m} \nonumber \\
	&+ \mc{O}(\Delta\bsla_{ji}^4) \ , \nonumber
\end{align}
where $\partial_{\lambda_j^k}g(\x,\bsla_i,\bsla_j) \equiv \frac{\partial}{\partial_{\lambda_j^k}}g(\x,\bsla_i,\bsla_j)$ is the partial derivative of $g(\x,\bsla_i,\bsla_j)$ with respect to the $k$th component of the $\bsla$ vector, and $[\cdots]_{\lambda_{i}^k}$ indicates that the argument is evaluated at $\lambda_{j}^k = \lambda_i^k$.  We have also made use of Einstein summation notation, where repeated indices are implicitly summed over.

The first RHS term is
\begin{equation}
	g(\x,\bsla_i,\bsla_i) = \pi_i(\x)\ln\frac{\pi_i(\x)}{\pi_i(\x)} = \pi_i\ln 1 = 0 \label{eq:app:TSS-work-expansion-term1} \ .
\end{equation}
The RHS first-derivative term in \eqref{eq:app:TSS-work-expansion-2} is
\begin{subequations}
\begin{align}
\partial&_{\lambda_j^k}g(\x,\bsla_i,\bsla_j) \nonumber\\
&= \left( \partial_{\lambda_j^k}\sqrt{\pi_i\pi_j} \right)\ln\frac{\pi_i}{\pi_j} +\sqrt{\pi_i\pi_j}\left( \partial_{\lambda_j^k}\ln\frac{\pi_i}{\pi_j} \right) \label{eq:app:TSS-work-expansion-3-1}\\
&= \frac{1}{2}\left( \partial_{\lambda_j^k}\pi_j \right)\sqrt{\frac{\pi_i}{\pi_j}}\ln\frac{\pi_i}{\pi_j} - \sqrt{\pi_i\pi_j}\frac{1}{\pi_j}\left(\partial_{\lambda_j^k}\pi_j\right) \label{eq:app:TSS-work-expansion-3-2}
\end{align}\label{eq:app:TSS-work-expansion-3}
\end{subequations}
which when evaluated at $\lambda_j^k = \lambda_i^k$ gives
\begin{subequations}
\begin{align}
\left[ \partial_{\lambda_j^k}g(\x,\bsla_i,\bsla_j) \right]_{\lambda_i^k} &= \frac{1}{2}\left[ \partial_{\lambda_j^k}\pi_j \right]_{\lambda_i^k}\ln 1 - \left[ \partial_{\lambda_j^k}\pi_j \right]_{\lambda_i^k} \label{eq:app:TSS-work-expansion-4-1}\\
&= -\left[ \partial_{\lambda_j^k}\pi_j \right]_{\lambda_i^k} \label{eq:app:TSS-work-expansion-4-2}
\end{align}\label{eq:app:TSS-work-expansion-4}\ .
\end{subequations}

Substituting \eqref{eq:app:TSS-work-expansion-3} into the RHS second-derivative term of \eqref{eq:app:TSS-work-expansion-2} gives
\begin{align}
	\partial_{\lambda_j^k}\partial_{\lambda_j^{\ell}}&\left[ \sqrt{\pi_i\pi_j}\ln\frac{\pi_i}{\pi_j} \right] \label{eq:app:TSS-work-expansion-5} \\
	&= \partial_{\lambda_j^{\ell}}\left[ \frac{1}{2}\left( \partial_{\lambda_j^k}\pi_j \right)\sqrt{\frac{\pi_i}{\pi_j}}\ln\frac{\pi_i}{\pi_j} \right. \nonumber \left. - \sqrt{\frac{\pi_i}{\pi_j}}\left( \partial_{\lambda_j^k}\pi_j \right) \right] \ . \nonumber
\end{align}	
The second RHS term,
\begin{subequations}
\label{eq:app:TSS-work-expansion-6}
\begin{align}
	\partial&_{\lambda_j^{\ell}}\left[ \sqrt{\frac{\pi_i}{\pi_j}}\left( \partial_{\lambda_j^k}\pi_j \right) \right] \nonumber \\
	&= \left( \partial_{\lambda_j^k}\sqrt{\frac{\pi_i}{\pi_j}} \right)\left( \partial_{\lambda_j^k}\pi_j \right) + \sqrt{\frac{\pi_i}{\pi_j}}\left( \partial^2_{\lambda_j^{\ell},\lambda_j^k} \pi_j \right) \label{eq:app:TSS-work-expansion-6-1} \\
	&= -\frac{1}{2}\sqrt{\frac{\pi_i}{\pi_j^3}}\left( \partial_{\lambda_j^{\ell}}\pi_j \right)\left( \partial_{\lambda_j^k}\pi_j \right) + \sqrt{\frac{\pi_i}{\pi_j}}\left( \partial^2_{\lambda_j^{\ell}\lambda_j^k}\pi_j \right) \label{eq:app:TSS-work-expansion-6-2} \ ,
\end{align}
\end{subequations}
evaluated at $\lambda_j^k=\lambda_i^k$ and $\lambda_j^{\ell}=\lambda_i^{\ell}$ gives
\begin{align}
\partial_{\lambda_j^{\ell}}&\left[ \sqrt{\frac{\pi_i}{\pi_j}}\left( \partial_{\lambda_j^k}\pi_j \right) \right] \label{eq:app:TSS-work-expansion-7} \\
&= -\frac{1}{2\pi_i}\left[ \partial_{\lambda_j^{\ell}}\pi_j \right]_{\lambda_i^{\ell}}\left[ \partial_{\lambda_j^k} \pi_j \right]_{\lambda_i^k} + \left[ \partial^2_{\lambda_j^{\ell}\lambda_j^k} \pi_j \right]_{\lambda_i^k\lambda_i^{\ell}} \ . \nonumber
\end{align}
The first RHS term of \eqref{eq:app:TSS-work-expansion-5},
\begin{align}
\partial_{\lambda_j^{\ell}}&\left[ \frac{1}{2}\left( \partial_{\lambda_j^k}\pi_j \right)\sqrt{\frac{\pi_i}{\pi_j}}\ln\frac{\pi_i}{\pi_j} \right] = \label{eq:app:TSS-work-expansion-8}\\
&\frac{1}{2}\sqrt{\frac{\pi_i}{\pi_j}}\ln\left(\frac{\pi_i}{\pi_j}\right)\left( \partial^2_{\lambda_j^{\ell}\lambda_j^k} \pi_j \right) \nonumber - \frac{1}{2}\left( \partial_{\lambda_j^{\ell}}\pi_j \right)\left( \partial_{\lambda_j^k}\pi_j \right)\sqrt{\frac{\pi_i}{\pi_j^3}} \nonumber\\ 
&\quad - \frac{1}{4}\left(\partial_{\lambda_j^{\ell}}\pi_j \right)\left( \partial_{\lambda_j^k}\pi_j \right)\sqrt{\frac{\pi_i}{\pi_j^3}}\ln\frac{\pi_i}{\pi_j} \nonumber
\end{align}	
evaluated at $\lambda_j^k=\lambda_i^k$ and $\lambda_j^{\ell}=\lambda_i^{\ell}$ gives
\begin{subequations}
\label{eq:app:TSS-work-expansion-9}
\begin{align}
\partial&_{\lambda_j^{\ell}}\left[ \frac{1}{2}\left( \partial_{\lambda_j^k}\pi_j \right)\sqrt{\frac{\pi_i}{\pi_j}}\ln\frac{\pi_i}{\pi_j} \right]_{\lambda_i^k\lambda_i^{\ell}} \nonumber \\
&=\frac{1}{2}\left[ \partial^2_{\lambda_j^{\ell}\lambda_j^k}\pi_j \right]_{\lambda_i^{\ell}\lambda_i^k}\ln 1 - \frac{1}{2\pi_i}\left[\partial_{\lambda_j^k}\pi_i \right]_{\lambda_i^k}\left[ \partial_{\lambda_j^{\ell}}\pi_j \right]_{\lambda_i^{\ell}} \nonumber \\
&\quad - \frac{1}{4\pi_i}\left[\partial_{\lambda_j^k}\pi_i \right]_{\lambda_i^k}\left[ \partial_{\lambda_j^{\ell}}\pi_j \right]_{\lambda_i^{\ell}}\ln 1 \label{eq:app:TSS-work-expansion-9-1}\\
&= -\frac{1}{2\pi_i}\left[\left( \partial_{\lambda_j^k}\pi_j \right)\left( \partial_{\lambda_j^{\ell}}\pi_j \right)\right]_{\lambda_i^k\lambda_i^{\ell}} \label{eq:app:TSS-work-expansion-9-2} \ .
\end{align}
\end{subequations}

Substituting \eqref{eq:app:TSS-work-expansion-7} and \eqref{eq:app:TSS-work-expansion-9} into \eqref{eq:app:TSS-work-expansion-5} gives
\begin{equation}
\partial^2_{\lambda_j^{\ell}\lambda_j^{k}}\left[ \sqrt{\pi_i\pi_j}\ln\frac{\pi_i}{\pi_j} \right]_{\lambda_i^{\ell}\lambda_i^k} = -\left[\partial^2_{\lambda_j^{\ell}\lambda_j^k}\pi_j \right]_{\lambda_i^{\ell}\lambda_j^k} \label{eq:app:TSS-work-expansion-10} \ .
\end{equation}
Substituting \eqref{eq:app:TSS-work-expansion-10}, \eqref{eq:app:TSS-work-expansion-4}, and \eqref{eq:app:TSS-work-expansion-term1} into \eqref{eq:app:TSS-work-expansion-2} gives the $\mc{O}(\Delta\bsla^2)$ approximation of $g(\x,\bsla_i,\bsla_j)$:
\begin{align}
	g(\x,\bsla_i,\bsla_j) \approx &-\left[ \partial_{\lambda_j^k}\pi_j \right]_{\lambda_i^k}\Delta\lambda_{ji}^k \label{eq:app:TSS-work-expansion-11} \\
	&- \frac{1}{2}\left[ \partial^2_{\lambda_j^{\ell}\lambda_j^k}\pi_j \right]_{\lambda_i^{\ell}\lambda_i^k}\Delta\lambda_{ji}^k\Delta\lambda_{ji}^{\ell} + \mc{O}(\Delta\bsla^3) \nonumber \ .
\end{align}
Both derivatives of $\pi_j$ vanish upon integration over $\x$:
\begin{subequations}
\label{eq:app:TSS-work-integral-zero}
\begin{align}
	\int\partial_{\lambda_j^k}\pi_j\, \md \x &= \partial_{\lambda_j^k}\int \pi_j\, \md \x = 0 \\
	\int \partial^2_{\lambda_j^{\ell}\lambda_j^k}\pi_j\, \md \x &= \partial_{\lambda_j^{\ell}} \partial_{\lambda_j^k}
	\int \pi_j\, \md\x = 0 \ , 
\end{align}
\end{subequations}
indicating that $\langle\beta W_{\rm ex}^{\rm TSS}\rangle_{ij}$ is third order in $\Delta\bsla$: in contrast to the discrete-control TSS excess work in \cite{large_2019}, the second-order term vanishes.

Substituting \eqref{eq:app:TSS-work-expansion-6} and \eqref{eq:app:TSS-work-expansion-8} into \eqref{eq:app:TSS-work-expansion-5} and differentiating gives the third-derivative term in \eqref{eq:app:TSS-work-expansion-2}: 
\begin{align}
	&\partial^3_{\lambda_j^m\lambda_j^{\ell}\lambda_j^k}\bigg[ \sqrt{\pi_i\pi_j}\left.\ln\frac{\pi_i}{\pi_j} \right] = \label{eq:app:TSS-work-expansion-12} \\ &\partial_{\lambda_j^m}\left[ \left( \partial^2_{\lambda_j^{\ell}\lambda_j^k}\pi_j \right)\sqrt{\frac{\pi_i}{\pi_j}}\phi_{ji} - \frac{1}{4}\left( \partial_{\lambda_j^{\ell}}\pi_j \right)\left(\partial_{\lambda_j^k}\pi_j \right)\sqrt{\frac{\pi_i}{\pi_j^3}} \ln\frac{\pi_i}{\pi_j} \right] \nonumber \ .
\end{align}
We have defined
\begin{equation}
	\phi_{ji} \equiv \frac{1}{2}\ln\frac{\pi_i}{\pi_j} - 1 \label{eq:app:TSS-work-phi} \ ,
\end{equation}
such that
\begin{equation}
	\partial_{\lambda_j^m}\phi_{ji} = -\frac{1}{\pi_j}\left(\partial_{\lambda_j^m}\pi_j \right)\label{eq:app:TSS-work-phi-deriv} \ ,
\end{equation}
and at $\bsla_j=\bsla_i$, $\phi_{ji}=-1$ and $\partial_{\lambda_j^m}\phi_{ji}=-\frac{1}{2\pi_i}[\partial_{\lambda_j^m}\pi_j]_{\lambda_i^m}$.

Thus the first RHS term of \eqref{eq:app:TSS-work-expansion-12} is
\begin{subequations}
\begin{align}
	\partial&_{\lambda_j^m}\left[ \left( \partial_{\lambda_j^{\ell}\lambda_j^k}^2 \pi_j \right) \sqrt{\frac{\pi_i}{\pi_j}} \phi_{ji} \right] \nonumber \\
	&= \left( \partial_{\lambda_j^m\lambda_j^{\ell}\lambda_j^k}^3 \pi_j \right)\sqrt{\frac{\pi_i}{\pi_j}}\phi_{ji} + \left( \partial_{\lambda_j^{\ell}\lambda_j^k}^2 \pi_j \right)\left( \partial_{\lambda_j^m}\phi_{ji}\right)\sqrt{\frac{\pi_i}{\pi_j}} \nonumber \\
	&\quad - \frac{1}{2}\left( \partial_{\lambda_j^{\ell}\lambda_j^k}^2\pi_j \right)\left( \partial_{\lambda_j^m}\pi_j \right)\sqrt{\frac{\pi_i}{\pi_j^3}}\phi_{ji} \label{eq:app:TSS-work-expansion-13} \\
	&= -\left[ \partial_{\lambda_j^m\lambda_j^{\ell}\lambda_j^k}^3 \pi_j \right]_{\lambda_i^m\lambda_i^{\ell}\lambda_i^k} \label{eq:app:TSS-work-expansion-14}\ , \quad \bsla_j=\bsla_i \ .
\end{align}
\end{subequations}
The second RHS term of \eqref{eq:app:TSS-work-expansion-12} is
\begin{subequations}
\begin{align}
	\partial_{\lambda_j^m}&\bigg[ \left( \partial_{\lambda_j^{\ell}}\pi_j \right)\left( \partial_{\lambda_j^k}\pi_j \right)\sqrt{\frac{\pi_i}{\pi_j^3}}\ln\frac{\pi_i}{\pi_j} \bigg] \nonumber \\ 
	&= \left( \partial_{\lambda_j^m\lambda_j^{\ell}}^2\pi_j \right)\left( \partial_{\lambda_j^k}\pi_j \right)\sqrt{\frac{\pi_i}{\pi_j^3}}\ln\frac{\pi_i}{\pi_j} \nonumber \\
	&\quad + \left( \partial_{\lambda_j^{\ell}\pi_j}\pi_j\right)\left( \partial_{\lambda_j^m\lambda_j^k}^2\pi_j \right)\sqrt{\frac{\pi_i}{\pi_j^3}}\ln\frac{\pi_i}{\pi_j} \label{eq:app:TSS-work-expansion-15} \\
	&\quad -\frac{1}{3}\left( \partial_{\lambda_j^m}\pi_j \right)\left( \partial_{\lambda_j^{\ell}}\pi_j \right)\left( \partial_{\lambda_j^k}\pi_i\right)\sqrt{\frac{\pi_i}{\pi_j^5}}\ln\frac{\pi_i}{\pi_j} \nonumber \\
	&\quad -\left( \partial_{\lambda_j^m}\pi_j \right)\left( \partial_{\lambda_j^{\ell}}\pi_j \right)\left( \partial_{\lambda_j^k}\pi_i \right)\sqrt{\frac{\pi_i}{\pi_j^5}} \nonumber \\
	&= -\frac{1}{\pi_i^2}\left[ \partial_{\lambda_j^{m}}\pi_j \right]_{\lambda_j^m}\left[ \partial_{\lambda_j^{\ell}} \pi_j \right]_{\lambda_j^{\ell}}\left[ \partial_{\lambda_j^k}\pi_j \right]_{\lambda_j^k} \ , \quad \bsla_j=\bsla_i \ . 
	\label{eq:app:TSS-work-expansion-16} 
\end{align}
\end{subequations}

Substituting \eqref{eq:app:TSS-work-expansion-16} and \eqref{eq:app:TSS-work-expansion-14} into \eqref{eq:app:TSS-work-expansion-12} gives
\begin{align}
	\partial_{\lambda_j^m\lambda_j^{\ell}\lambda_j^k}^3&\left[ \sqrt{\pi_i\pi_j}\ln\frac{\pi_i}{\pi_j} \right] = -\left[ \partial_{\lambda_j^m\lambda_j^{\ell}\lambda_j^k}^3 \pi_j \right]_{\lambda_i^m\lambda_i^{\ell}\lambda_i^k} \label{eq:app:TSS-work-expansion-17} \\ 
	& + \frac{1}{4\pi_i^2} \left[\partial_{\lambda_j^m}\pi_j \right]_{\lambda_j^m}\left[ \partial_{\lambda_j^{\ell}}\pi_j \right]_{\lambda_j^{\ell}}\left[ \partial_{\lambda_j^k}\pi_j \right]_{\lambda_j^k} \nonumber \ .
\end{align}
The first RHS term vanishes upon integration for the same reasons as \eqref{eq:app:TSS-work-integral-zero}. Thus, the leading-order approximation of \eqref{eq:app:TSS-work-expansion-1} in the small-$\Delta\bsla$ limit is
\begin{align}
	&\langle\beta W_{\rm ex}^{\rm TSS}\rangle_{ji} \approx  \label{eq:app:TSS-work-expansion-18} \\ &\frac{\Delta\lambda_{ji}^m\Delta\lambda_{ji}^{\ell}\Delta\lambda_{ji}^k}{4!}\int\frac{1}{\pi_i^2}\left[ \left(\partial_{\lambda_j^m}\pi_j\right) \left(\partial_{\lambda_j^{\ell}}\pi_j\right) \left(\partial_{\lambda_j^k}\pi_j\right) \right]_{\bsla_i}\md x \ . \nonumber
\end{align}

Using $\partial_{\lambda_j^k}\ln\pi_j = \frac{1}{\pi_j}\partial_{\lambda_j^k}\pi_j$ (and evaluating the resulting expression at $\bsla_j = \bsla_i$) simplifies the integral to the expectation of three derivatives of the log-probability,
\begin{subequations}
\label{eq:app:TSS-work-expansion-flexion-1}
\begin{align}
	&\langle\beta W_{\rm ex}^{\rm TSS}\rangle_{ji} \nonumber \\
	&\approx \frac{\Delta\lambda_{ji}^m\Delta\lambda_{ji}^{\ell}\Delta\lambda_{ji}^k}{4!} \left\langle \left(\partial_{\lambda_j^m}\ln\pi_j\right)\left(\partial_{\lambda_j^{\ell}}\ln\pi_j \right)\left( \partial_{\lambda_j^k}\ln\pi_j \right) \right\rangle_{\bsla_i} \label{eq:app:TSS-work-expansion-flexion-1-1}\\
	&=\frac{\Delta\lambda_{ji}^m\Delta\lambda_{ji}^{\ell}\Delta\lambda_{ji}^k}{4!} \mc{S}_{m\ell k}(\bsla_i) \label{eq:app:TSS-work-expansion-flexion-2} \ ,
\end{align}
\end{subequations}
where third-rank tensor $\mc{S}_{m\ell k}(\bsla)$ is the leading-order non-Gaussian approximation of the log-probability in the limit of small $\Delta\bsla$~\cite{selletin_2014}. $\mc{S}_{m\ell k}$ is related to the \emph{flexion tensor} used in the analysis of astrophysical image data. The name `flexion tensor' derives from the original use of flexion as a measure of third-order distortions in astrophysical images due to weak gravitational lensing~\cite{bacon_2006,goldberg_2005}.  

For a physical system in contact with a heat reservoir, the derivative of the log-probability is
\begin{subequations}
\label{eq:app:TSS-work-expansion-19}
\begin{align}
	\partial_{\lambda^k}\ln\pi &= \beta\left( f_k|_{\bsla} + \partial_{\lambda^k} F(\bsla) \right) \label{eq:app:TSS-work-expansion-19-1}\\
	&= \delta f_k|_{\bsla} \label{eq:app:TSS-work-expansion-19-2}
\end{align}
\end{subequations}
where $f_{k}|_{\bsla} \equiv -\partial_{\lambda^k}E(x|\bsla)|_{\bsla}$ is the generalized force conjugate to control parameter $\lambda^k$ at $\bsla$, and $F(\bsla) = \langle E\rangle_{\bsla} - TS = -\lambda^k\langle f_k\rangle_{\bsla} - TS$ is the equilibrium free energy at $\bsla$.  

Substituting \eqref{eq:app:TSS-work-expansion-19} into the TSS excess work~\eqref{eq:app:TSS-work-expansion-flexion-2} gives
\begin{equation}
	\langle\beta W_{\rm ex}^{\rm TSS}\rangle_{ji} \approx \frac{1}{4!}\Delta\lambda_{ji}^m\Delta\lambda_{ji}^{\ell}\Delta\lambda_{ji}^k\langle\delta f_m\delta f_{\ell}\delta f_k\rangle_{\bsla_i} \label{eq:app:TSS-work-expansion-flexion-3} \ ,
\end{equation}
for third centered moment $\langle \delta f_m\delta f_{\ell}\delta f_k\rangle_{\bsla_i}$ of the generalized forces at chemical state $\bsla_i$.

This analysis shows two primary features: in the TSS limit the excess work due to an autonomous subsystem discretely transitioning between states can be calculated exactly through~(8b), and in the small-$\Delta\bsla$ limit, the leading-order contribution to the TSS excess work~\eqref{eq:app:TSS-work-expansion-flexion-2} is $\mc{O}(\Delta\bsla^3)$. Furthermore, for physical systems in contact with thermal reservoirs, the third-rank flexion tensor can be expressed as a matrix of third centered moments of the conjugate forces~\eqref{eq:app:TSS-work-expansion-flexion-3}.

\section{\label{suppsec:neq-work}Nonequilibrium excess work in autonomous systems}

Here we consider the additional excess work in a system driven out of equilibrium by biased chemical dynamics. In particular, we write the total excess work associated with a particular chemical transition $\bsla_i\to\bsla_j$ as the sum of two components
\begin{equation}
	\langle\beta W_{\rm ex}\rangle_{ji} = \langle\beta W_{\rm ex}^{\rm TSS}\rangle_{ji} + \langle\beta W_{\rm ex}^{\rm neq}\rangle_{ji} \ ,
\end{equation}
where $\langle\beta W_{\rm ex}^{\rm TSS}\rangle_{ji}$ is the excess work in the timescale-separated limit (Sec.~\ref{suppsec:TSS-power}), and $\langle\beta W_{\rm ex}^{\rm neq}\rangle_{ji}$ is the nonequilibrium excess work, the additional excess work required of the chemical dynamics due to the mechanical subsystem being out of equilibrium.  

In order to give a general form for the nonequilibrium excess work, we appeal to linear-response theory. Specifically, we evaluate the integral expression
\begin{align}
	\langle &W\rangle_{ji} = \label{eq:app:neq-work-integral-1} \\
	&\int \left[ E(\x|\bsla_j) - E(\x|\bsla_i) \right]p_{\rm neq}(\x,t)p^{\rm dwell}_{ji}(\x,t)\, \md \x \, \md t \nonumber \ ,
\end{align}
the total work done on the mechanical subsystem by the chemical dynamics for a transition $\bsla_i\to\bsla_j$, averaged over mechanical states $\x$ and times $t$. $p_{\rm neq}(\x,t)$ is the nonequilibrium distribution over mechanical states $\x$ at time $t$, and $p^{\rm dwell}_{ji}(\x,t)$ is 
the dwell-time distribution for the $\bsla_i\to\bsla_j$ transition while in mechanical state $\x$. We assume that at the microstate level, the chemical jump dynamics are Markovian with dwell-time distribution
\begin{equation}
	p^{\rm dwell}_{ji}(\x,t) = \W{ji}^{\x} e^{-\W{ji}^{\x}t} \label{eq:app:neq-work-dwell-dist} \ ,
\end{equation}
with rates $\W{ji}^{\x}$ given in (3).

We make a weak-perturbation approximation, Taylor expanding the energy landscape $E(\x|\bsla_j)$ around $E(\x|\bsla_i)$:
\begin{subequations}
\label{eq:app:linear-energy-expansion}
\begin{align}
	E(\x|\bsla_j) &\approx E(\x|\bsla_i) + \left[\nabla_{\bsla}\cdot E(\x|\bsla)\right]_{\bsla_i} \left(\bsla_j - \bsla_i \right) \label{eq:app:linear-energy-expansion-1} \\
	&= E(\x|\bsla_i) - f_k|_{\bsla_i}\Delta\lambda^k_{ji} \ . \label{eq:app:linear-energy-expansion-2}
\end{align}
\end{subequations}
This simplifies the rates~(3) to
\begin{subequations}
\label{eq:app:rate-simplification}
\begin{align}
	\W{ji}^{\x} &= \Gamma_{ji}\, e^{-\frac{1}{2}\beta\left[ E(\x|\bsla_j) - E(\x|\bsla_i) \right]} \label{eq:app:rate-simplification-1} \\
	&\approx \Gamma_{ji}\, e^{\frac{1}{2}\beta f_k|_{\bsla_i}\Delta\lambda^k_{ji}} \label{eq:app:rate-simplification-2} \\
	&\approx \Gamma_{ji}\, \left( 1 + \tfrac{1}{2}\beta f_k|_{\bsla_i}\Delta\lambda^k_{ji} \right) \ , \label{eq:app:rate-simplification-3}
\end{align}
\end{subequations}
where $\Gamma_{ji} \equiv \Gamma\exp\left\{-\tfrac{1}{2}\beta\Delta\mu_{ji}\right\}$.  Substituting \eqref{eq:app:neq-work-dwell-dist}, (\ref{eq:app:rate-simplification}b,c), and \eqref{eq:app:linear-energy-expansion} simplifies the excess work~\eqref{eq:app:neq-work-integral-1} to 
\vspace{0.1in}

\begin{widetext}
\begin{subequations}
\label{eq:app:neq-work-integral-2}
\begin{align}	
	\langle W\rangle_{ji} &\approx -\Gamma_{ji}\Delta\lambda^k_{ji}\int f_k|_{\bsla_i}(\x)p_{\rm neq}(\x,t)e^{-\frac{1}{2}\beta\Delta E_{ji}}e^{-\Gamma_{ji}\left( 1 + \frac{1}{2}\beta\Delta E_{ji} \right) t}\, \md \x \, \md t \label{eq:app:neq-work-integral-2-1} \\
	&\approx -\Gamma_{ji}\Delta\lambda^k_{ji}\int_0^{\infty} e^{-\Gamma_{ji}t}\left\{ \sum_{n=0}^{\infty} \frac{(-1)^n}{n!}\left(\Gamma_{ji}t\right)^n \int f_k|_{\bsla_{i}}(\x)\left(\tfrac{1}{2}\beta\Delta E_{ji}\right)^n p_{\rm neq}(\x,t)e^{\frac{1}{2}\beta\Delta E_{ji}}\, \md \x  \right\} \md t \label{eq:app:neq-work-integral-2-2} \ ,
\end{align}
\end{subequations}
\end{widetext}
where $\Delta E_{ji} \equiv f_k|_{\bsla_i}(\x)\Delta\lambda^k_{ji}$. The second line Taylor expands $\exp\{-\Gamma_{ji}\tfrac{1}{2}\beta\Delta E_{ji} t\}$ about $t=0$. 

We now simplify \eqref{eq:app:neq-work-integral-2} using linear-response theory, effectively assuming that the most recent chemical transition is the dominant contribution to the present mechanical distribution~\cite{large_2019}. The true `initial' distribution over mechanical states following the previous chemical transition $\bsla_{\ell} \to \bsla_i$ is the (nonequilibrium) switching-position distribution $p_{i\ell}^{\rm sw}(\x,t)$. We approximate the difference between the mean conjugate force during relaxation from the previous nonequilibrium switching-position distribution $p_{i\ell}^{\rm sw}(\x,t)$ and at the (next) nonequilibrium switching-position distribution $p_{ji}^{\rm sw}(\x,t)$ (that enters into the exact nonequilibrium excess work $\langle\beta W_{\rm ex}^{\rm neq}\rangle_{ji}$), by the difference between the mean conjugate force during relaxation from the previous equilibrium distribution $\pi_{\ell}(\x)$ and at the current equilibrium distribution $\pi_i(\x)$. Due to the symmetries of the model, these two pairs of distributions should differ by similar amounts. This substitution of one mean excess conjugate force for another is accurate when the conjugate forces $f_{\x|\bsla_i}$ are approximately linear in $\x$ for all $\x$ with significant probability in $p_{\rm neq}(\x,t)$, $p_{ji}^{\rm sw}(\x,t)$, and the intervening relaxation. This approximation is trivially satisfied for a harmonic confining potential (such as in ``Linear-transport motor'' section of the main text) 
where conjugate forces are always linear, and approximately satisfied for more general cases in the small-$\Delta\bsla$ limit.

First-order Taylor expanding the mechanical potential $E(\x|\bsla_{\ell})$ around $\bsla_{\ell}=\bsla_i$ (similar to \eqref{eq:app:linear-energy-expansion}) gives an approximate form for the previous equilibrium distribution
\begin{equation}
    \pi_{\ell}(\x) = \frac{1}{\mc{Z}_{\ell}}\pi_i(\x) e^{-\beta\Delta E_{i\ell}} \ .
\end{equation}
Thus, using $\pi_{\ell}(\x)$ as the initial condition for $p_{\rm neq}(\x,t)$ at $t=0$, we use linear-response theory to approximate the spatial integral in the $n$th term of \eqref{eq:app:neq-work-integral-2-2} in terms of correlation functions~\cite{chandler}:
\begin{subequations}
\label{eq:app:neq-work-integral-3}
\begin{align}
	I_k^{(n)}(t) &\equiv
	\int g_{\bsla_i}^{(n)}(\x) p_{\rm neq}(\x,t)e^{-\frac{1}{2} \beta\Delta E_{ji}} \, \md \x \\
	&= \frac{1}{\mc{Z}_{\ell}}
	\int g_{\bsla_i}^{(n)}(\x)
	\pi_i(\x) \, e^{-\beta\Delta E_{i\ell}} e^{\frac{1}{2}\beta\Delta E_{ji}} \, \md \x  \label{eq:app:neq-work-integral-3-1}\\
	&\approx \left\langle g_{\bsla_i}^{(n)}\right\rangle_{\bsla_i} - \left\langle g_{\bsla_i}^{(n)}(t)\beta\Delta E_{i\ell}\right\rangle_{\bsla_i} \label{eq:app:neq-work-integral-3-2} \\ 
	&\quad + \left\langle g_{\bsla_i}^{(n)}\right\rangle_{\bsla_i}\langle \beta\Delta E_{i\ell}\rangle_{\bsla_i} + \frac{1}{2}\left\langle g_{\bsla_i}^{(n)}(t)\beta\Delta E_{ji} \right\rangle_{\bsla_i} \nonumber \ .
\end{align}
\end{subequations}
Here the angle brackets $\langle\cdots\rangle_{\bsla_i}$ indicate an average over the equilibrium distribution $\pi_i(\x)$, and $g_{\bsla_i}^{(n)}(\x) \equiv f_k|_{\bsla_i}(\x)\left( \beta\Delta E_{ji} \right)^n$.

To calculate the nonequilibrium excess work beyond the TSS work~\eqref{eq:app:tss-work} outlined in Sec.~\ref{suppsec:TSS-power}, we subtract the linear-response approximation to the TSS work from \eqref{eq:app:neq-work-integral-3-2}. The approximate TSS work (including the free energy change), within the linear-response approximation of $g_{\bsla}^{(n)}(\x)$ averaged over the equilibrium switching-position distribution $p^{\rm sw}_{ji}(\x)$, is
\begin{subequations}
\begin{align}
	\langle g_{\bsla_i}^{(n)}\rangle^{\rm sw}_{ji} &= \frac{1}{\mc{Z}^{\rm sw}_{ji}}\int g_{\bsla_i}^{(n)}(\x) p^{\rm sw}_{ji}(\x)\, \md \x \\
	&\approx \frac{1}{\mc{Z}^{\rm sw}_{ji}}\int g_{\bsla_{i}}^{(n)}(\x) \pi_i(\x) \left( 1 + \tfrac{1}{2}\beta f_k\Delta\lambda_{ji}^k \right) \md \x \\
	&= \left\langle g_{\bsla_i}^{(n)}\right\rangle_{\bsla_i} + \frac{1}{2}\left\langle g_{\bsla_i}^{(n)}(t)\beta\Delta E_{ji}\right\rangle_{\bsla_i} \ ,
\end{align}
\end{subequations}
which simplifies $I_k^{(n)}(t)$ to
\begin{align}
	I_k^{(n)}(t) &= \left\langle g_{\bsla_i}^{(n)}\right\rangle^{\rm sw}_{ji} \label{eq:app:neq-work-integral-4} \\
	&\quad - \beta \left[\left\langle g_{\bsla_i}^{(n)}(t)\Delta E_{i\ell}\right\rangle_{\bsla_i} + \left\langle g_{\bsla_i}^{(n)}\right\rangle_{\bsla_i}\langle\Delta E_{i\ell}\rangle_{\bsla_i}\right] \nonumber \ .
\end{align}

The first RHS term approximates the TSS work, so subtracting $\langle g_{\bsla_i}^{(n)}\rangle^{\rm sw}_{ji}$ from both sides gives
\begin{subequations}
\begin{align}
	&\delta I_k^{(n)}(t) \equiv I_k^{(n)}(t) - \left\langle g_{\bsla_i}^{(n)}\right\rangle^{\rm sw}_{ji} \\
	&= - \frac{\beta^{n+1}}{2^n} \Delta\lambda_{i\ell}^{k'}\left(\Delta\lambda_{ji}^{k'}\right)^{n}\langle\delta f_k^{n+1}(t)\delta f_{k'}(0)\rangle_{\bsla_i} \label{eq:app:neq-work-integral-5} \ .
\end{align}
\end{subequations}
Beyond the leading-order ($n=0$) contribution, every term is $\mc{O}(\Delta\bsla^2)$ or higher, hence beyond the approximation made in \eqref{eq:app:neq-work-integral-4}.  

In \eqref{eq:app:neq-work-integral-2}, when the relaxation rates of the force autocovariances $\langle\delta f_k(0)\delta f_{k'}(t)\rangle_{\bsla_i}$ are significantly larger than $\Gamma_{ji}$, the Taylor series is well-approximated by its leading term:
\begin{align}
	\sum_{n=0}^{\infty}&\frac{(-1)^n}{n!}\left( \Gamma_{ji} t \right)^n\int g_{\bsla_i}^{(n)}(t)p_{\rm neq}(\x,t)e^{-\frac{1}{2}\beta\Delta E_{ji}} \, \md \x \nonumber \\
	&\approx - \beta\Delta\lambda_{i\ell}^k\langle\delta f_k(t)\delta f_{k'}(0)\rangle_{\bsla_i} + \left\langle g_{\bsla_i}^{(0)}\right\rangle_{ji}^{\rm sw} \ . \label{eq:app:neq-work-integral-6}
\end{align}
This approximation amounts to a statement of timescale separation between the chemical and mechanical dynamics, when the mechanical degrees of freedom relax significantly faster than the chemical-state dynamics.  For instance, if the force autocovariance decays exponentially, $\langle\delta f_k(t)\delta f_{k'}(0)\rangle_{\bsla_i} \propto \exp\{-k_{\rm relax}t\}$, then this approximation~\eqref{eq:app:neq-work-integral-5} holds when $k_{\rm relax} \gg \Gamma_{ji}$. 

Substituting \eqref{eq:app:neq-work-integral-6} in the work~\eqref{eq:app:neq-work-integral-2} and subtracting the linear-response approximation to the TSS work $\langle g_{\bsla_i}^{(0)}\rangle_{\bsla_i}$ from both sides, gives the nonequilibrium excess work
\begin{align}
	\langle W_{\rm ex}^{\rm neq}\rangle_{ji|\ell} &\approx  \label{eq:app:neq-work-integral-7} \\ &\beta\Gamma_{ji}\Delta\lambda_{ji}^{k}\Delta\lambda_{i\ell}^{k'}\int_0^{\infty}\langle\delta f_k(t)\delta f_{k'}(0)\rangle_{\bsla_i}e^{-\Gamma_{ji}t}\, \md t \nonumber \ ,
\end{align}
which is the excess work required for the $\bsla_i\to\bsla_j$ transition (given that chemical state $\bsla_{\ell}$ immediately precedes $\bsla_i$), beyond the work required in the TSS limit. The average $\langle\cdots\rangle_{ji|\ell}$ is conditioned on the previous chemical state $\bsla_{\ell}$.  Once again, we expand the exponential as $\exp\{-\Gamma_{ji}t\} \approx 1$ (based on the same approximation simplifying the Taylor series in \eqref{eq:app:neq-work-integral-6}), simplifying the nonequilibrium excess work~\eqref{eq:app:neq-work-integral-7} to
\begin{subequations}
\label{eq:app:neq-excess-work-1}
\begin{align}	
	\langle W_{\rm ex}^{\rm neq}\rangle_{ji|\ell} &\approx \Gamma_{ji} \Delta\lambda_{ji}^k\Delta\lambda_{i\ell}^{k'}\beta\int_0^{\infty}\langle\delta f_{k}(t)\delta f_{k'}(0)\rangle_{\bsla_i}\md t \label{eq:app:neq-excess-work-1-1}\\
	&= \Gamma_{ji} \Delta\lambda_{ji}^k\Delta\lambda_{i\ell}^{k'}\zeta_{kk'}(\bsla_i) \label{eq:app:neq-excess-work-1-2}
\end{align}
\end{subequations}
where $\zeta_{kk'}(\bsla) \equiv \beta\int_0^{\infty}\langle\delta f_k(0)\delta f_{k'}(t)\rangle_{\bsla}\md t$ is the \emph{generalized friction tensor} originally derived for continuous, deterministic control~\cite{sivak_2012}.

This approximation depends on three chemical states: $\bsla_{\ell}$, $\bsla_i$, and $\bsla_j$. To eliminate the dependence on $\bsla_{\ell}$, we average over $\bsla_{\ell}$, weighted by $P_{\ell}V_{i\ell}/V_{i*}$, the coarse-grained transition rate of $\bsla_{\ell}\to\bsla_i$ divided by the total transition rate $V_{i*} \equiv \sum_{s} P_{s}V_{is}$ into state $\bsla_i$.  Thus within the linear-response regime, the average excess work for transition $\bsla_i\to\bsla_j$, averaged over dwell-time fluctuations and previous states, is
\begin{equation}
	\langle\beta W_{\rm ex}^{\rm neq}\rangle_{ji}\approx \frac{\beta\Gamma_{ji}}{V_{i*}} \Delta\lambda_{ji}^k\zeta_{kk'}(\bsla_i)\sum_{s}P_s V_{is}\Delta\lambda_{is}^{k'} \label{eq:app:neq-excess-work-2} \ .
\end{equation}

The corresponding approximate nonequilibrium excess power is
\begin{subequations}
\label{eq:app:neq-excess-work-3}
\begin{align}
	\langle&\beta\mc{P}_{\rm ex}^{\rm neq}\rangle_{\Lambda\to X} \approx \sum_{ji}P_i \langle \beta\mc{P}_{\rm ex}^{\rm neq}\rangle_{ji} \label{eq:app:neq-excess-work-3-1} \\
	&= \sum_{ji} P_iV_{ji}\langle\beta W_{\rm ex}^{\rm neq}\rangle_{ji} \label{eq:app:neq-excess-work-3-2} \\
	&\approx \sum_{ji}P_i \frac{\beta V_{ji}\Gamma_{ji}}{V_{i*}} \Delta\lambda_{ji}^{k}\zeta_{kk'}(\bsla_i)\sum_{s}P_sV_{is}\Delta\lambda_{is}^{k'} \label{eq:app:neq-excess-work-3-3} \ .
\end{align}
\end{subequations} 
All terms in \eqref{eq:app:neq-excess-work-3} can be determined from coarse-grained observations, aside from the generalized friction tensor $\zeta_{kk'}(\bsla)$, a phenomenological quantity determined from conditional equilibrium measurements of the mechanical degrees of freedom. 

Equation~\eqref{eq:app:neq-excess-work-3} can be greatly simplified in particular scenarios. For instance, we consider a system with a single chemical coordinate (one-dimensional $\lambda$) with uniform step sizes $\Delta\lambda$, a uniform generalized friction $\zeta$, uniform forward and reverse coarse-grained rates $V_{\pm}$, and equal coarse-grained probabilities ($P_{i}=P$) for all $i$). The average excess work for a forward ($+$) or reverse ($-$) step is
\begin{equation}
	\langle\beta W_{\rm ex}^{\rm neq}\rangle_{\pm} \approx \pm \beta\zeta\Gamma_{\pm}\Delta\lambda^2\frac{V_+ - V_-}{V_+ + V_-} \label{eq:app:neq-work-1D-1} \ .
\end{equation}
Here, $\Gamma_{\pm} \equiv \Gamma\exp\{\pm\tfrac{1}{2}\beta\Delta\mu\}$, for chemical potential difference $\Delta\mu$ for a forward chemical step. The $\pm$ prefactor in \eqref{eq:app:neq-work-1D-1} reflects that, on average, forward steps require positive work input while reverse steps require negative work input. Thus the average work done for any chemical transition (forward or reverse) is the mean of $\langle\beta W_{\rm ex}^{\rm neq}\rangle_{+}$ and $\langle\beta W_{\rm ex}^{\rm neq}\rangle_{-}$, each term weighted by the corresponding $\pm$ jump probabilities:
\begin{subequations}
\label{eq:app:neq-work-1D-2}
\begin{align}
	\langle&\beta W_{\rm ex}^{\rm neq}\rangle_{\Delta\lambda} \approx \frac{V_+\langle\beta W_{\rm ex}^{\rm neq}\rangle_{+}}{V_+ + V_-} + \frac{V_-\langle\beta W_{\rm ex}^{\rm neq}\rangle_{-}}{V_+ + V_-} \label{eq:app:neq-work-1D-2-1}\\
	&= \beta\zeta\Delta\lambda^2 \frac{V_+ - V_-}{(V_+ + V_-)^2} \left( V_+\Gamma_+ - V_-\Gamma_{-} \right) \label{eq:app:neq-work-1D-2-2} \ .
\end{align}
\end{subequations}

\section{\label{suppsec:simulation-details}Simulation details}

In the harmonic model system (``Linear-transport motor'' in main text), 
the mechanical coordinate is simulated using an overdamped Langevin equation:
\begin{equation}
	\md x = -\beta D\partial_x E(x|\lambda) \, \md t + \sqrt{2D} \, \md W(t) \ , \label{eq:app:langevin} 
\end{equation}
where $D$ is the diffusion coefficient and $W(t)$ is a standard Wiener process. The chemical transitions are governed by the coarse-grained discrete master equation
\begin{equation}
\dot{P}_j = \sum_{i}P_{i}V_{ji} \ ,
\end{equation}
where $P_i = \int p(x,t|\lambda_i) \, \md x$ is the coarse-grained probability of chemical state $\lambda_i$ and 
\begin{equation}
V_{ji} = \int \W{ji}(x)p(x,t|\lambda_{i}) \, \md x \label{eq:app:CG-rate}
\end{equation}
is the coarse-grained rate of chemical transition $\lambda_i\to\lambda_j$. For a system at steady state, the coarse-grained rates~\eqref{eq:app:CG-rate} become time-independent.

We implement the joint mechanochemical dynamics using kinetic Monte Carlo~\cite{bataille_2008}, where the probability of a chemical transition $\lambda_i\to\lambda_j$ occurring in a given time step $\Delta t$ is determined by the current system state $x_t$ and the transition rate $W_{ji}(x_t)$. Specifically, the probability of a transition $\bsla_i\to\bsla_j$ occurring for a system in state $x_t$ in a time interval between time $t$ and $t+\Delta t$ is $1 - \exp\{-\W{ji}(x_t)\Delta t\}\Delta t$. 

In the model for a linear-transport motor in the main text, the nonequilibrium excess work (14) equals the sum of changes $\Delta E_{ji} = E(x_t|\lambda_j) - E(x_t|\lambda_i)$ in mechanical state energy during chemical transitions. In particular, for a trajectory with $N_{\Lambda}$ chemical transitions $\lambda_0\to\lambda_1\to\cdots\to\lambda_{N_{\Lambda}}$, with each transition occurring at times $t_{\lambda_{10}},t_{\lambda_{21}},\cdots,t_{\lambda_{N_{\Lambda},N_{\Lambda-1}}}$, the average excess work per step is 
\begin{subequations}
\label{eq:app:sim-work-heat}
\begin{align}
	\langle\beta &W_{\rm ex}^{\rm neq}\rangle_{\Delta\lambda} = \frac{\beta W}{N_{\Lambda}} \\
	&= \frac{\beta}{N_{\Lambda}} \sum_{i=0}^{N_{\Lambda}-1}E(x_{t_{i+1,i}}|\lambda_{i+1}) - E(x_{t_{i+1,i}}|\lambda_i) \label{eq:app:sim-work} \ .
\end{align}
\end{subequations}
The work equals the nonequilibrium excess work in \eqref{eq:app:sim-work} because, for this system the equilibrium free energy change is zero ($\Delta F_{ji} = 0$) and $\langle\beta W_{\rm ex}^{\rm TSS}\rangle_{ji} = 0$. 

\end{document}